\documentclass[submission,copyright,creativecommons]{eptcs}
\providecommand{\event}{ACL2 Workshop 2022}
\usepackage{underscore, latexsym,pifont,amsmath,amstext,amssymb,amsfonts,makeidx,url}
\title{A Formalization of Finite Group Theory}
\author{David M.~Russinoff
\email{david@russinoff.com}
}
\def\titlerunning{A Formalization of Finite Group Theory}
\def\authorrunning{D.M. Russinoff}

\begin{document}
\maketitle

\begin{abstract}
Previous formulations of group theory in ACL2 and Nqthm, based on either {\tt encapsulate} or {\tt defn\-sk}, have been limited by their
failure to provide a path to proof by induction on the order of a group, which is required for most interesting results in this domain beyond
Lagrange's Theorem (asserting the divisibility of the order of a group by that of a subgroup).  We describe an alternative approach to finite group
theory that remedies this deficiency, based on an explicit representation of a group as an operation table.  We define a {\tt defgroup} macro
for generating parametrized families of groups, which we apply to the additive and multiplicative groups of integers modulo $n$, the symmetric
groups, arbitrary quotient groups, and cyclic subgroups.  In addition to a proof of Lagrange's Theorem, we provide an inductive proof of
a theorem of Cauchy: {\it If the order of a group $G$ is divisible by a prime $p$, then $G$ has an element of order $p$.}
\end{abstract}

\section{Introduction}

Since ACL2 provides only limited support for quantification, modeling group theory in its logic is a challenging problem.  A 1990 paper of Yuan
Yu \cite{yu} presents a formal development of finite group theory in Nqthm based on the {\tt defn-sk} macro (surviving in ACL2 as {\tt defun-sk}),
which he uses to define a predicate that characterizes a list whose members satisfy the group axioms with respect to a fixed operation.  He
also defines the notion of group homomorphism and proves that the kernel of any homomorphism is a normal subgroup, but this requires an additional
{\tt defn-sk} form in order to introduce a group with a different operation. The culmination of Yu's work is Lagrange's Theorem: {\it The order
of a finite group is divisible by the order of any subgroup.}  Any significant further development of the theory would require induction on
the order of a group, which seems to be inaccessible through this method.

We are unaware of any ACL2 results in this domain that duplicate Yu's achievement.  A similar approach based on {\tt encapsulate} has been
suggested \cite{eric}, but while this provides a generalization to infinite groups, it otherwise shares the same limitations as the {\tt defun-sk}
method.  Heras et al.~\cite{heras} describe "a guideline to develop tools that simplify the formalizations related to algebraic structures in ACL2",
but do not mention any results that are directly relevant to the theory of groups.

We shall present an ACL2 formalization of finite groups that provides for inductive proofs as well as computations on concrete groups.  Our
scheme is based on the definition of a group as an explicit operation table, i.e., a matrix of group elements.  We define a {\tt defgroup} macro
that provides definitions of parametrized families of groups, which we apply to the additive and multiplicative groups of integers modulo $n$, the
symmetric and alternating groups, arbitrary quotient groups, and cyclic subgroups. Our proof of Lagrange's theorem shares some features with Yu's,
but we prove a stronger version stating that the order of a group is the product of that of a subgroup and its index.  This leads naturally into an
analysis of quotient groups, which lays the groundwork for a theorem of Cauchy: {\it If the order of a group $G$ is divisible by a prime $p$, then
$G$ has an element of order $p$.}  We present an inductive proof of this result, which illustrates the effectuality of our scheme.  The proof, which
resides in {\tt books/workshops/2022/russinoff-groups/}, uses several number-theoretic results from
{\tt books/projects/quadratic-reciprocity/euclid.lisp}, including the following basic theorem of Euclid:
{\it If a product of integers $ab$ is divisible by a prime $p$, then $p$ divides either $a$ or $b$.}

\section{Groups and Subgroups}\label{grps}

In our formalization, a group is a square matrix, the first row of which is a list of the group elements:
\begin{small}
\begin{verbatim}
  (defmacro elts (g) `(car ,g))
  (defmacro in (x g) `(member-equal ,x (elts ,g)))
  (defmacro order (g) `(len (elts ,g)))
\end{verbatim}
\end{small}
The {\it index} of a group element is its position in the list {\tt (elts g)}:
\begin{small}
\begin{verbatim}
  (defun index (x l)
    (if (consp l)
        (if (equal x (car l))
            0
          (1+ (index x (cdr l))))
      ()))
  (defmacro ind (x g) `(index x (elts g)))
\end{verbatim}
\end{small}
The group operation is defined as a table access:
\begin{small}
\begin{verbatim}
  (defun op (x y g) (nth (ind y g) (nth (ind x g) g)))
\end{verbatim}
\end{small}
We also define
\begin{small}
\begin{verbatim}
  (defmacro e (g) `(car (elts ,g)))
\end{verbatim}
\end{small}
Note that {\tt (nth (ind (e g) g) g)} = {\tt (nth 0 g)} = {\tt (elts g)} and therefore,
\begin{center}
  {\tt (op (e g) y g)} = {\tt (nth (ind y g) (elts g))} = {\tt y}
\end{center}
i.e., {\tt (e g)} is a left identity:
\begin{small}
\begin{verbatim}
  (defthm group-left-identity
    (implies (in x g) (equal (op (e g) x g) x)))
\end{verbatim}
\end{small}
The left inverse {\tt (inv x g)}, if it exists, is the group element of least index satisfying
\begin{center}
  {\tt (op (inv x g) x g)} = {\tt x}
\end{center}
defined as follows:
\begin{small}
\begin{verbatim}
  (defun inv-aux (x l g)
    (if (consp l)
        (if (equal (op (car l) x g) (e g))
            (car l)
          (inv-aux x (cdr l) g))
      ()))
  (defun inv (x g) (inv-aux x (elts g) g))
\end{verbatim}
\end{small}
The definition of a group is based on a set of predicates, including those representing the group axioms:
\begin{small}
\begin{verbatim}
  (defund groupp (g)
    (and (posp (order g))
         (matrixp g (order g) (order g))
         (dlistp (elts g))
         (not (in () g))
         (closedp g)
         (assocp g)
         (inversesp g)))
\end{verbatim}
\end{small}
The predicate {\tt matrixp} is a straightforward characterization of a matrix of given dimensions, and {\tt dlistp} recognizes lists
of distinct members. The condition that {\tt NIL} is not a group element is unnecessary but avoids certain technical difficulties.
The definition of {\tt closedp} recursively searches for a a pair of group elements {\tt x} and {\tt y} such that {\tt (op x y g)}
is not in the group.  This allows us to prove the theorem
\begin{small}
\begin{verbatim}
  (defthm group-closure
    (implies (and (groupp g) (in x g) (in y g))
             (in (op x y g) g)))
\end{verbatim}
\end{small}
and also provides a counterexample if the search succeeds:
\begin{small}
\begin{verbatim}
  (defthm not-closedp-cex
    (implies (and (dlistp (elts g)) (not (closedp g)))
             (let* ((cex (closedp-cex g)) (x (car cex)) (y (cadr cex)))
               (and (in x g) (in y g)
                    (not (in (op x y g) g))))))
\end{verbatim}
\end{small}
The latter result is useful in verifying {\tt (closedp g)} for a conjectured group {\tt g}.
An analogous pair of theorems is derived for {\tt assocp}, which searches for a triple that violates associativity, and {\tt inversesp},
which searches for an element without a left inverse.   Other basic properties, e.g., right identity, right inverse, and cancellation laws,
follow easily.

Similarly, subgroups are defined by the predicate
\begin{small}
\begin{verbatim}
  (defun subgroupp (h g)
    (and (groupp g)
         (groupp h)
         (sublistp (elts h) (elts g))
         (not (subgroupp-cex h g))))
\end{verbatim}
\end{small}
where {\tt subgroupp-cex} exhaustively searches for a pair of elements of {\tt h} on which the two group operations disagree.  Thus,
\begin{small}
\begin{verbatim}
  (defthm subgroup-op
    (implies (and (subgroupp h g) (in x h) (in y h))
             (equal (op x y h) (op x y g))))
\end{verbatim}
\end{small}
Again, the search produces a counter-example if it exists:
\begin{small}
\begin{verbatim}
  (defthm not-subgroupp-cex
    (implies (and (groupp g) (groupp h)
                  (sublistp (elts h) (elts g))
                  (not (subgroupp h g)))
             (let* ((cex (subgroupp-cex h g)) (x (car cex)) (y (cadr cex)))
               (and (in x h) (in y h)
                    (not (equal (op x y h) (op x y g)))))))
\end{verbatim}
\end{small}
The following results are also readily derived from the definition:
\begin{small}
\begin{verbatim}
  (defthm subgroup-e
    (implies (subgroupp h g) (equal (e h) (e g))))

  (defthm subgroup-inv
    (implies (and (subgroupp h g) (in x h))
             (equal (inv x h) (inv x g))))
\end{verbatim}
\end{small}
We also define a function {\tt subgroup}, which constructs the subgroup of a given group with a given element list if such a group exists.
An illustration will be provided in the next section.

\section{Parametrized Groups}

The macro {\tt defgroup} is based on the following encapsulation, which constrains three functions representing the list of elements of a
group, the group operation, and its inverse operator:
\begin{small}
\begin{verbatim}
  (encapsulate (((glist) => *) ((gop * *) => *) ((ginv *) => *))
    (local (defun glist () (list 0)))
    (local (defun gop (x y) (+ x y)))
    (local (defun ginv (x) x))
    (defthm consp-glist (consp (glist)))
    (defthm dlistp-glist (dlistp (glist)))
    (defthm g-non-nil (not (member-equal () (glist))))
    (defthm g-identity
      (implies (member-equal x (glist))
               (equal (gop (car (glist)) x) x)))
    (defthm g-closed
      (implies (and (member-equal x (glist))
                    (member-equal y (glist)))
               (member-equal (gop x y) (glist))))
    (defthm g-assoc
      (implies (and (member-equal x (glist))
                    (member-equal y (glist))
                    (member-equal z (glist)))
               (equal (gop x (gop y z)) (gop (gop x y) z))))
    (defthm g-inverse
      (implies (member-equal x (glist))
               (and (member-equal (ginv x) (glist))
                    (equal (gop (ginv x) x) (car (glist)))))))
\end{verbatim}
\end{small}
Our definition of the group {\tt (g)} is based on the constrained functions {\tt gop} and {\tt glist}:
\begin{small}
\begin{verbatim}
  (defun g-row (x m)
    (if (consp m)
        (cons (gop x (car m)) (g-row x (cdr m)))
      ()))
  (defun g-aux (l m)
    (if (consp l)
        (cons (g-row (car l) m) (g-aux (cdr l) m))
      ()))
  (defun g () (let ((l (glist))) (g-aux l l)))
\end{verbatim}
\end{small}
Using the results discussed in Section~\ref{grps}, we prove the theorem
\begin{small}
\begin{verbatim}
  (defthm groupp-g (groupp (g)))
\end{verbatim}
\end{small}
along with three results characterizing the group structure:
\begin{small}
\begin{verbatim}
  (defthm glist-elts (equal (elts (g)) (glist)))
  (defthm op-g-rewrite
    (implies (and (in x (g)) (in y (g)))
             (equal (op x y (g)) (gop x y))))
  (defthmd inv-g-rewrite (implies (in x (g)) (equal (inv x (g)) (ginv x))))
\end{verbatim}
\end{small}
The macro defines a parametrized family of groups given a parameter list, a predicate that the parameters must satisfy, and three terms
corresponding to the above constrained functions.  For example, the additive group of integers modulo {\tt n} is generated by the following:
\begin{small}
\begin{verbatim}
  (defun ninit (n) (if (zp n) () (append (ninit (1- n)) (list (1- n)))))
  (defun z+-op (x y n) (mod (+ x y) n))
  (defun z+-inv (x n) (mod (- x) n))
  (defgroup z+ (n) (posp n) (ninit n) (z+-op x y) (z+-inv x))
\end{verbatim}
\end{small}
Prior to the evaluation of a {\tt defgroup} form, a set of preliminary rewrite rules corresponding to the seven exported theorems of the
encapsulation must be proved.  The first three, which state that the specified list of elements is a non-{\tt NIL} list of distinct non-{\tt NIL}
members, are generally trivial.  In this case, the remaining four are also easy to prove:
\begin{small}
\begin{verbatim}
  (defthm z+-identity
    (implies (and (posp n) (member-equal x (ninit n)))
             (equal (z+-op 0 x n) x)))
  (defthm z+-closed
    (implies (and (posp n)
                  (member-equal x (ninit n))
                  (member-equal y (ninit n)))
             (member-equal (z+-op x y n) (ninit n))))
  (defthm z+-assoc
    (implies (and (posp n)
                  (member-equal x (ninit n))
                  (member-equal y (ninit n))
                  (member-equal z (ninit n)))
             (equal (z+-op x (z+-op y z n) n)
                    (z+-op (z+-op x y n) z n))))
  (defthm z+-inverse
    (implies (and (posp n) (member-equal x (ninit n)))
             (and (member-equal (z+-inv x n) (ninit n))
                  (equal (z+-op (z+-inv x n) x n) 0))))
\end{verbatim}
\end{small}
The family {\tt (z+ n)} is then defined by the above {\tt defgroup} form, which also automatically proves four theorems:
\begin{small}
\begin{verbatim}
  (DEFTHM GROUPP-Z+ (IMPLIES (POSP N) (GROUPP (Z+ N))))
  (DEFTHM Z+-ELTS (IMPLIES (POSP N) (EQUAL (ELTS (Z+ N)) (NINIT N))))
  (DEFTHM Z+-OP-REWRITE
    (IMPLIES (AND (POSP N) (IN X (Z+ N)) (IN Y (Z+ N)))
             (EQUAL (OP X Y (Z+ N)) (Z+-OP X Y N))))
  (DEFTHM Z+-INV-REWRITE
    (IMPLIES (AND (POSP N) (IN X (Z+ N)))
             (EQUAL (INV X (Z+ N)) (Z+-INV X N))))
\end{verbatim}
\end{small}
Each of these results is derived by the same functional instantiation of the corresponding lemma pertaining to the constrained constant {\tt g}.
For example,
\begin{small}
\begin{verbatim}
  (DEFTHM GROUPP-Z+ (IMPLIES (POSP N) (GROUPP (Z+ N)))
    :HINTS (("Goal"
             :USE ((:FUNCTIONAL-INSTANCE GROUPP-G
                     (GLIST (LAMBDA NIL (IF (POSP N) (NINIT N) (GLIST))))
                     (GOP (LAMBDA (X Y) (IF (POSP N) (Z+-OP X Y) (GOP X Y))))
                     (GINV (LAMBDA (X) (IF (POSP N) (Z+-INV X) (GINV X))))
                     (G-ROW (LAMBDA (X M) (IF (POSP N) (Z+-ROW X M N) (G-ROW X M))))
                     (G-AUX (LAMBDA (L M) (IF (POSP N) (Z+-AUX L M N) (G-AUX L M))))
                     (G (LAMBDA NIL (IF (POSP N) (Z+ N) (G)))))))))
\end{verbatim}
\end{small}
Note that {\tt Z+-ROW} and {\tt Z+-AUX} are auxiliary functions that are generated by {\tt defgroup} along with {\tt Z+}.

The multiplicative group {\tt (z* n)} of integers modulo {\tt n} is similarly generated, replacing addition with multiplication:
\begin{small}
\begin{verbatim}
  (defun z*-op (x y n) (mod (* x y) n))
\end{verbatim}
\end{small}
where we assume {\tt (and (natp n) (> n 1))}.  The element list is the sublist {\tt (rel-primes n)} of {\tt (ninit n)} consisting of integers
relatively prime to {\tt n}.  This list is computed using the greatest common divisor function, {\tt g-c-d}, which is treated in {\tt euclid.lisp}.
It is clear that {\tt (car (rel-primes n))} = 1 satisfies the identity property.  Closure and associativity follow from the established
properties of {\tt g-c-d} and {\tt mod}.  For the definition of the inverse operator, we appeal to the following property of {\tt g-c-d}: 
\begin{small}
\begin{verbatim}
  (defthm g-c-d-linear-combination
    (implies (and (integerp x) (integerp y))
             (= (+ (* (r-int x y) x) (* (s-int x y) y))
                (g-c-d x y))))
\end{verbatim}
\end{small}
Thus, {\tt (g-c-d x y)} is a linear combination of {\tt x} and {\tt y} with integer coefficients {\tt (r-int x y)} and {\tt (s-int x y)}.
We define
\begin{small}
\begin{verbatim}
  (defun z*-inv (x n) (mod (r-int x n) n))
\end{verbatim}
\end{small}
and the required property follows from {\tt g-c-d-linear-combination}:
\begin{small}
\begin{verbatim}
  (defthm z*-inverse
    (implies (and (natp n) (> n 1) (member-equal x (rel-primes n)))
             (and (member-equal (z*-inv x n) (rel-primes n))
                  (equal (z*-op (z*-inv x n) x n) 1))))
\end{verbatim}
\end{small}
Thus, we have
\begin{small}
\begin{verbatim}
  (defgroup z* (n) (and (natp n) (> n 1)) (rel-primes n) (z*-op x y n) (z*-inv x n))
\end{verbatim}
\end{small}
and the usual four generated theorems, including
\begin{small}
\begin{verbatim}
  (DEFTHM GROUPP-Z* (IMPLIES (AND (NATP N) (> N 1)) (GROUPP (Z* N))))
\end{verbatim}
\end{small}
For example, evaluation of {\tt (z* 15)} yields a group or order 8:
\begin{small}
\begin{verbatim}
  ((1 2 4 7 8 11 13 14)
   (2 4 8 14 1 7 11 13)
   (4 8 1 13 2 14 7 11)
   (7 14 13 4 11 2 1 8)
   (8 1 2 11 4 13 14 7)
   (11 7 14 2 13 1 8 4)
   (13 11 7 1 14 8 4 2)
   (14 13 11 8 7 4 2 1))
\end{verbatim}
\end{small}
As an illustration of the {\tt subgroup} function mentioned at the end of Section~\ref{grps}, we observe that
\begin{small}
\begin{verbatim}
  (subgroup '(1 4 7 13) (z* 15))
\end{verbatim}
\end{small}
is
\begin{small}
\begin{verbatim}
  ((1 4 7 13)
   (4 1 13 7)
   (7 13 4 1)
   (13 7 1 4))
\end{verbatim}
\end{small}
and {\tt (subgroupp (subgroup '(1 4 7 13) (z* 15)) (z* 15))} = {\tt T}. Note that in order for {\tt sub\-group} to succeed in generating
a group, the first member of the supplied list must be the identity element.\medskip

The element list of the symmetric group {\tt (sym n)} is given by
\begin{center}
  {\tt (defund slist (n) (perms (ninit n)))}
\end{center}
where {\tt (perms l)} returns a list of all permutations of a list {\tt l}.  The group operation is composition, defined by
\begin{small}
\begin{verbatim}
  (defun comp-perm-aux (p r l)
    (if (consp l)
        (cons (nth (nth (car l) r) p) (comp-perm-aux p r (cdr l)))
      ()))
  (defun comp-perm (p r n) (comp-perm-aux p r (ninit n)))
\end{verbatim}
\end{small}
and the inverse operator is
\begin{small}
\begin{verbatim}
  (defun inv-perm-aux (p l)
    (if (consp l)
        (cons (index (car l) p) (inv-perm-aux p (cdr l)))
      ()))
  (defun inv-perm (p n) (inv-perm-aux p (ninit n)))
\end{verbatim}
\end{small}
Once we establish the required preliminary lemmas (which has not yet been done at the time of writing), we shall invoke
\begin{small}
\begin{verbatim}
  (defgroup sym (n) (posp n) (slist n) (comp-perm x y n) (inv-perm x n)).
\end{verbatim}
\end{small}
In the meantime, we have defined a weaker version of {\tt defgroup} that defines a family of groups without proving any theorems, and does not
require either the parameter constraint or the inverse operator:
\begin{small}
\begin{verbatim}
  (defgroup-light sym (n) (slist n) (comp-perm x y n))
\end{verbatim}
\end{small}
This allows us to analyze concrete groups of the family.  For example, {\tt (sym 3)} is a group of order 6,
\begin{small}
\begin{verbatim}
  (((0 1 2) (0 2 1) (1 0 2) (1 2 0) (2 0 1) (2 1 0))
   ((0 2 1) (0 1 2) (2 0 1) (2 1 0) (1 0 2) (1 2 0))
   ((1 0 2) (1 2 0) (0 1 2) (0 2 1) (2 1 0) (2 0 1))
   ((1 2 0) (1 0 2) (2 1 0) (2 0 1) (0 1 2) (0 2 1))
   ((2 0 1) (2 1 0) (0 2 1) (0 1 2) (1 2 0) (1 0 2))
   ((2 1 0) (2 0 1) (1 2 0) (1 0 2) (0 2 1) (0 1 2)))
\end{verbatim}
\end{small}
and we can prove
\begin{small}
\begin{verbatim}
  (defthm groupp-sym-3 (groupp (sym 3)))
\end{verbatim}
\end{small}
by direct computation.  Note that the identity element of {\tt (sym n)} is the trivial permutation {\tt (ninit n)}.

To construct the alternating groups, we define a function {\tt cyc} that converts an element of {\tt (sym n)} to an alternative representation
as a product of cycles.  For example, in {\tt (sym 5)}, 
\begin{center}
  {\tt (cyc '(2 3 4 1 0))} = {\tt ((0 2 4) (1 3))},
\end{center}
We can derive the parity of a permutation from the observation that a cycle of odd (resp., even) length is a product of an even (resp., odd) number of
transpositions.  Thus, we define an even permutation as follows:
\begin{small}
\begin{verbatim}
  (defun even-cyc (cyc)
    (if (consp cyc)
        (if (evenp (len (car cyc)))
            (not (even-cyc (cdr cyc)))
          (even-cyc (cdr cyc)))
      t))
  (defun even-perm (perm) (even-cyc (cyc perm)))
\end{verbatim}
\end{small}
The function {\tt even-perms} extracts the sublist of even permutations from a list.  The alternating group {\tt (alt n)} is the subgroup of
{\tt (sym n)} consisting of the even permutations:
\begin{small}
\begin{verbatim}
  (defgroup-light alt (n)
    (even-perms (slist n))
    (comp-perm x y n))  
\end{verbatim}
\end{small}
For example, {\tt (alt 3)} is the group
\begin{small}
\begin{verbatim}
  (((0 1 2) (1 2 0) (2 0 1))
   ((1 2 0) (2 0 1) (0 1 2))
   ((2 0 1) (0 1 2) (1 2 0)))
\end{verbatim}
\end{small}
and we can prove theorems such as {\tt (subgroupp (alt 5) (sym 5))}.

\section{Cosets and Lagrange's Theorem}

For our purposes, we need only consider left cosets.
Given an element {\tt x} and a subgroup {\tt h} of a group {\tt g}, {\tt (lcoset x h g)} is defined to be a list of all elements of {\tt g}
of the form {\tt (op x y g)} that satisfy {\tt (in y h)}.  In particular, {\tt (lcoset (e g) h g)} is a permutation of {\tt (elts h)}.  Our definition of
{\tt lcoset} ensures that this list is ordered by indices with respect to {\tt g}.  It follows that its members are distinct:
\begin{small}
\begin{verbatim}
  (defthm dlistp-lcosets
    (implies (and (subgroupp h g) (in x g))
             (dlistp (lcoset x h g)))))
\end{verbatim}
\end{small}
It is also easily shown that the length of each coset is the order of the subgroup:
\begin{small}
\begin{verbatim}
  (defthm len-lcoset
    (implies (and (subgroupp h g) (in x g))
             (equal (len (lcoset x h g)) (order h))))
\end{verbatim}
\end{small}
The following is a useful criterion for coset membership:
\begin{small}
\begin{verbatim}
  (defthmd member-lcoset-iff
    (implies (and (subgroupp h g) (in x g) (in y g))
             (iff (member-equal y (lcoset x h g))
                  (in (op (inv x g) y g) h))))
\end{verbatim}
\end{small}
As a consequence of this result, intersecting cosets have the same members, and the following may be derived from the ordering property:
\begin{small}
\begin{verbatim}
  (defthmd equal-lcoset
    (implies (and (subgroupp h g) (in x g) (in y g)
                  (member-equal y (lcoset x h g)))
             (equal (lcoset y h g) (lcoset x h g))))
\end{verbatim}
\end{small}

The list {\tt (lcosets h g)} is constructed by traversing {\tt (elts g)} and adding a coset to the list whenever an element is encountered
that does not already appear in the list.  By definition, the length of the list is the index of {\tt h} in {\tt g}:
\begin{small}
\begin{verbatim}
  (defun subgroup-index (h g) (len (lcosets h g)))
\end{verbatim}
\end{small}
Our proof of Lagrange's Theorem is based on the list {\tt (append-list (lcosets h g))}, 
produced by appending all members of {\tt (lcosets h g)}.  The above results lead to the following properties of this list:
\begin{small}
\begin{verbatim}
  (defthm dlistp-append-list-lcosets
    (implies (subgroupp h g)
             (dlistp (append-list (lcosets h g)))))
  (defthm len-lcosets
    (implies (subgroupp h g)
             (equal (len (append-list (lcosets h g)))
                    (* (order h) (subgroup-index h g)))))
\end{verbatim}
\end{small}
The proof of Lagrange's Theorem depends on the observation that if each of two lists of distinct members is a sublist of the other, then the
lists have the same length:
\begin{small}
\begin{verbatim}
  (defthmd sublistp-equal-len
    (implies (and (dlistp l)
                  (dlistp m)
                  (sublistp l m)
                  (sublistp m l))
             (equal (len l) (len m))))
\end{verbatim}
\end{small}
The hypotheses of this lemma are readily established for the lists {\tt (append-list (lcosets h g))} and {\tt (elts g)}, and the theorem follows:
\begin{small}
\begin{verbatim}
  (defthm lagrange
    (implies (and (groupp g) (subgroupp h g))
             (equal (* (order h) (subgroup-index h g))
                    (order g))))
\end{verbatim}
\end{small}

\section{Normal Subgroups and Quotient Groups}\label{normal}

For elements {\tt x} and {\tt y} of a group {\tt g}, the conjugate of {\tt x} by {\tt y} is defined by
\begin{small}
\begin{verbatim}
  (defund conj (x y g) (op (op (inv y g) x g) y g))
\end{verbatim}
\end{small}
Note that this computation returns {\tt x} iff {\tt x} and {\tt y} commute.

A normal subgroup {\tt h} of {\tt g} is recognized by the predicate {\tt (normalp h g)}, which first requires that {\tt h} be a subgroup of {\tt g}
and then exhaustively checks that every conjugate of every element of {\tt h} is an element of {\tt h}.  As usual, we have the following
two results:
\begin{small}
\begin{verbatim}
  (defthm normalp-conj
    (implies (and (normalp h g) (in x h) (in y g))
             (in (conj x y g) h)))
  (defthmd not-normalp-cex
    (let* ((cex (normalp-cex h g)) (x (car cex)) (y (cadr cex)))
      (implies (and (subgroupp h g) (not (normalp h g)))
               (and (in x h) (in y g) (not (in (conj x y g) h))))))
\end{verbatim}
\end{small}

We shall apply {\tt defgroup} to define the group {\tt (quotient g h)} when {\tt h} is a normal subgroup of {\tt g}.
The elements of this group are the members of {\tt (lcosets h g)}, and the identity element is the coset of {\tt (e g)}:
\begin{small}
\begin{verbatim}
  (defun qe (h g) (lcoset (e g) h g))
\end{verbatim}
\end{small}
Thus, we must rearrange {\tt (lcosets h g)}, moving this element to the front of the list:
\begin{small}
\begin{verbatim}
  (defun qlist (h g) (cons (qe h g) (remove1-equal (qe h g) (lcosets h g))))
\end{verbatim}
\end{small}
The group operation is
\begin{small}
\begin{verbatim}
  (defun qop (x y h g) (lcoset (op (car x) (car y) g) h g))
\end{verbatim}
\end{small}
and the inverse operator is
\begin{small}
\begin{verbatim}
  (defun qinv (x h g) (lcoset (inv (car x) g) h g))
\end{verbatim}
\end{small}
The closure property is trivial.  The remaining properties required by {\tt defgroup} (identity, associativity, and inverse) may be derived
from the following result, which is a consequence of {\tt normalp-conj} and {\tt member-lcoset-iff}:
\begin{small}
\begin{verbatim}
(defthm op-qop
  (implies (and (normalp h g)
                (member-equal x (qlist h g)) (member-equal y (qlist h g))
                (member-equal a x) (member-equal b y))
           (member-equal (op a b g) (qop x y h g))))
\end{verbatim}
\end{small}
We may now invoke
\begin{small}
\begin{verbatim}
  (defgroup quotient (g h) (normalp h g) (qlist h g) (qop x y h g) (qinv x h g))
\end{verbatim}
\end{small}
which generates the usual four results, including
\begin{small}
\begin{verbatim}
  (DEFTHM GROUPP-QUOTIENT (IMPLIES (NORMALP H G) (GROUPP (QUOTIENT H G))))
\end{verbatim}
\end{small}
It is easily shown that any subgroup of index 2 is normal.  For example, by direct computation,
\begin{center}
  {\tt (normalp (alt 5) (sym 5))} = {\tt T}.
\end{center}
As another example, the element {\tt (1 2 0)} of {\tt (sym 3)} generates a subgroup of order 3 in a group of order 6, and therefore,
\begin{center}
  {\tt (normalp (subgroup '((0 1 2) (1 2 0) (2 0 1)) (sym 3)))} = {\tt T}.
\end{center}
 According to {\tt GROUPP-QUOTIENT}, its quotient group
\begin{small}
\begin{verbatim}
  (quotient (sym 3) (subgroup '((0 1 2) (1 2 0) (2 0 1)) (sym 3)))
\end{verbatim}
\end{small}
 is a group of order 2:
\begin{small}
\begin{verbatim}
  ((((0 1 2) (1 2 0) (2 0 1)) ((0 2 1) (1 0 2) (2 1 0)))
   (((0 2 1) (1 0 2) (2 1 0)) ((0 1 2) (1 2 0) (2 0 1))))
\end{verbatim}
\end{small}
Of course, any subgroup of an abelian group is normal.  For example, {\tt (subgroup '(1 3 9) (z* 13))} is a normal subgroup of {\tt (z* 13)} of
index 4.  Its quotient group is
\begin{small}
\begin{verbatim}
  (((1 3 9) (2 5 6) (7 8 11) (4 10 12))
   ((2 5 6) (4 10 12) (1 3 9) (7 8 11))
   ((7 8 11) (1 3 9) (4 10 12) (2 5 6))
   ((4 10 12) (7 8 11) (2 5 6) (1 3 9)))
\end{verbatim}
\end{small}

\section{Parametrized Subgroups}

The macro {\tt defsubgroup} calls {\tt defgroup} to define a subgroup of a given group {\tt g}.  The last two arguments of {\tt defgroup} are
not supplied to {\tt defsubgroup}, since they are always {\tt (op x y g)} and {\tt (inv x g)}.  As an illustration, the {\it centralizer} of an
element {\tt a} of {\tt g} is the subgroup consisting of all elements that commute with {\tt a}.  The definition of its element list,
{\tt (centizer-elts a g)}, is straightforward.  Several of the rewrite rules required by {\tt defgroup} are generated by {\tt defsubgroup},
but the following must be proved by the user:
\begin{small}
\begin{verbatim}
  (defthm dlistp-centizer-elts
    (implies (and (groupp g) (in a g)) (dlistp (centizer-elts a g))))
  (defthm sublistp-centizer-elts
    (implies (and (groupp g) (in a g)) (sublistp (centizer-elts a g) (elts g))))
  (defthm centizer-elts-identity
    (implies (and (groupp g) (in a g)) (equal (car (centizer-elts a g)) (e g))))
  (defthm consp-centizer-elts
    (implies (and (groupp g) (in a g)) (consp (centizer-elts a g))))
  (defthm centizer-elts-closed
    (implies (and (groupp g) (in a g) 
                  (member-equal x (centizer-elts a g))
                  (member-equal y (centizer-elts a g)))
             (member-equal (op x y g) (centizer-elts a g))))
  (defthm centizer-elts-inverse
    (implies (and (groupp g) (in a g) (member-equal x (centizer-elts a g)))
             (member-equal (inv x g) (centizer-elts a g))))
\end{verbatim}
\end{small}
we may then invoke
\begin{small}
\begin{verbatim}
(defsubgroup centralizer (a g) (and (groupp g) (in a g)) (centizer-elts a g))
\end{verbatim}
\end{small}
In addition to the lemmas generated by {\tt defsubgroup}, this produces
\begin{small}
\begin{verbatim}
  (DEFTHM SUBGROUPP-CENTRALIZER 
    (IMPLIES (AND (GROUPP G) (IN A G))
             (SUBGROUPP (CENTRALIZER A G) G)))
\end{verbatim}
\end{small}

The {\it center} of {\tt g} consists of all elements that commute with every element of {\tt g}.  The list of such elements, {\tt (cent-elts g)}
is again easily defined, and after proving the requisite rewrite rules, we have
\begin{small}
\begin{verbatim}
  (defsubgroup center (g) (groupp g) (cent-elts g))
\end{verbatim}
\end{small}

Our final example is the cyclic subgroup generated by an element {\tt a} of {\tt g}.  First we define the powers of {\tt a}:
\begin{small}
\begin{verbatim}
  (defun power (a n g)
    (if (zp n)
        (e g)
      (op a (power a (1- n) g) g)))
\end{verbatim}
\end{small}
The usual formulas for a product of powers and a power of a power are derived by induction:
\begin{small}
\begin{verbatim}
  (defthm power+
    (implies (and (groupp g) (in a g) (natp n) (natp m))
             (equal (op (power a n g) (power a m g) g)
                    (power a (+ n m) g))))
  (defthm power*
    (implies (and (groupp g) (in a g) (natp n) (natp m))
             (equal (power (power a n g) m g)
                    (power a (* n m) g))))
\end{verbatim}
\end{small}
Next, we define the order of {\tt a} in {\tt g}:
\begin{small}
\begin{verbatim}
  (defun ord-aux (a n g)
    (declare (xargs :measure (nfix (- (order g) n))))
    (if (equal (power a n g) (e g))
        n
      (if (and (natp n) (< n (order g)))
          (ord-aux a (1+ n) g)
        ())))
  (defun ord (a g) (ord-aux a 1 g))
\end{verbatim}
\end{small}
We cannot have {\tt (ord a g)} = {\tt NIL}, for it would then follow from {\tt power+} that the powers of {\tt a} include
$\mbox{\tt (order g)} + 1$ distinct elements.  This observation has the following consequences:
\begin{small}
\begin{verbatim}
  (defthm ord<=order
    (implies (and (groupp g) (in a g))
             (and (posp (ord a g)) (<= (ord a g) (order g)))))
  (defthm divides-ord
    (implies (and (groupp g) (in a g) (natp n))
             (iff (equal (power a n g) (e g)) (divides (ord a g) n))))
  (defthm power-mod
    (implies (and (groupp g) (in a g) (natp n))
             (equal (power a n g) (power a (mod n (ord a g)) g))))
  (defthm ord-power-div
    (implies (and (groupp g) (in a g) (posp n) (divides n (ord a g)))
             (equal (ord (power a n g) g) (/ (ord a g) n))))
\end{verbatim}
\end{small}
Thus, there are {\tt (ord a g)} distinct powers of {\tt a}. A list of these elements is computed by {\tt powers}:
\begin{small}
\begin{verbatim}
  (defun powers-aux (a n g)
    (if (zp n)
        ()
      (append (powers-aux a (1- n) g) (list (power a (1- n) g)))))
  (defun powers (a g) (powers-aux a (ord a g) g))
\end{verbatim}
\end{small}
The following are readily derived from the definition:
\begin{small}
\begin{verbatim}
  (defthm member-powers
    (implies (and (groupp g) (in a g)
                  (natp n) (< n (ord a g)))
             (equal (nth n (powers a g))
                    (power a n g))))
  (defthm power-index
    (implies (and (groupp g) (in a g)
                  (member-equal x (powers a g)))
             (equal (power a (index x (powers a g)))
                    x)))
\end{verbatim}
\end{small}
It follows from {\tt power-mod} that {\tt (powers a g)} is closed under the group operation, and it follows from {\tt power-index} and {\tt power+}
that for {\tt x} in {\tt (powers a g)},
\begin{small}
\begin{verbatim}
  (inv x g) = (power a (- (ord a g) (index x (powers a g))) g)
\end{verbatim}
\end{small}
and hence, {\tt (inv x g)} belongs to {\tt (powers a g)}.  The remaining prerequisites are trivial, and we have
\begin{small}
\begin{verbatim}
  (defsubgroup cyclic (a g)
    (and (groupp g) (in a g))
    (powers a g)
\end{verbatim}
\end{small}
Note that the two example subgroups at the end of Section~\ref{normal} can be computed as cyclic subgroups:
\begin{center}
  {\tt (subgroup '((0 1 2) (1 2 0) (2 0 1)) (sym 3))} = {\tt (cyclic '(1 2 0) (sym 3))}
\end{center}
and
\begin{center}
  {\tt (subgroup '(1 3 9) (z* 13))} = {\tt (cyclic 3 (z* 13))}.
\end{center}

As another example, the permutation {\tt (1 2 3 4 0)} of {\tt (sym 5)} is of order 5, and its cyclic subgroup
\begin{center}
  {\tt (cyclic '(1 2 3 4 0) (sym 5))}
\end{center}
is
\begin{small}
\begin{verbatim}
  (((0 1 2 3 4) (1 2 3 4 0) (2 3 4 0 1) (3 4 0 1 2) (4 0 1 2 3))
   ((1 2 3 4 0) (2 3 4 0 1) (3 4 0 1 2) (4 0 1 2 3) (0 1 2 3 4))
   ((2 3 4 9 1) (3 4 0 1 2) (4 0 1 2 3) (0 1 2 3 4) (1 2 3 4 0))
   ((3 4 0 1 2) (4 0 1 2 3) (0 1 2 3 4) (1 2 3 4 0) (2 3 4 0 1))
   ((4 0 1 2 3) (0 1 2 3 4) (1 2 3 4 0) (2 3 4 0 1) (3 4 0 1 2)))
\end{verbatim}
\end{small}

\section{Abelian Case of Cauchy's Theorem}

The formulation of Cauchy's Theorem requires a witness function, which searches a group for an element of a given order:
\begin{small}
\begin{verbatim}
  (defun elt-of-ord-aux (l p n)
    (if (consp l)
        (if (= (ord (car l) g) n)
            (car l)
          (elt-of-ord-aux (cdr l) n g))
      ()))
  (defun elt-of-ord (n g) (elt-of-ord-aux (elts g) n g))
\end{verbatim}
\end{small}
Thus, {\tt (elt-of-ord n g)} selects an element of {\tt g} of order {\tt n}, or returns {\tt NIL} if none exists:
\begin{small}
\begin{verbatim}
  (defthm elt-of-ord-ord
    (implies (and (groupp g) (natp n) (elt-of-ord n g))
             (and (in (elt-of-ord n g) g)
                  (equal (ord (elt-of-ord n g) g)
                         n))))
  (defthm elt-of-ord-ord
    (implies (and (groupp g) (natp n) (null (elt-of-ord n g)) (in a g))
             (not (= (ord a g) n))))
\end{verbatim}
\end{small}
Here are some simple examples:
\begin{itemize}
\item {\tt (elt-of-ord 5 (sym 5))} = {\tt (1 2 3 4 0)};
\item {\tt (elt-of-ord 22 (z* 23))} = 5, the least primitive root of 23;
\item {\tt (elt-of-ord (order (z* 35)) (z* 35))} = {\tt NIL}, since {\tt (z* 35)} is not cyclic.
\end{itemize}
The theorem may be stated as follows:
\begin{small}
\begin{verbatim}
  (defthm cauchy
    (implies (and (groupp g)
                  (primep p)
                  (divides p (order g)))
             (and (in (elt-of-ord p g) g)
                  (equal (ord (elt-of-ord p g) g) p))))
\end{verbatim}
\end{small}
Our proof, which closely adheres to the informal proof presented in \cite{rotman}, consists of two steps, both involving induction on the
order of {\tt g}.  First, it is proved for abelian {\tt g}, and then it is shown that if {\tt g} is nonabelian, then it must have a proper
subgroup with order divisible by {\tt p}.  Both steps depend critically on a result from {\tt euclid.lisp}:
\begin{small}
\begin{verbatim}
  (defthm euclid
      (implies (and (primep p)
                    (integerp a)
                    (integerp b)
                    (not (divides p a))
                    (not (divides p b)))
               (not (divides p (* a b)))))
\end{verbatim}
\end{small}
If {\tt g} has no element of order {\tt p}, then by {\tt ord-power-div}, it has no element of order divisible by {\tt p}, and
hence no cyclic subgroup of order divisible by {\tt p}.  Combining {\tt lagrange} and {\tt euclid}, we have
\begin{small}
\begin{verbatim}
  (defthm divides-order-quotient
    (implies (and (groupp g)
                  (primep p)
                  (divides p (order g))
                  (not (elt-of-ord p g))
                  (in a g))
             (divides p (order (quotient g (cyclic a g))))))
\end{verbatim}
\end{small}
By {\tt QUOTIENT-OP-REWRITE} and induction,
\begin{small}
\begin{verbatim}
  (defthm lcoset-power
    (implies (and (normalp h g) (in x g) (natp n))
             (equal (power (lcoset x h g) n (quotient g h))
                    (lcoset (power x n g) h g))))
\end{verbatim}
\end{small}
It follows that
\begin{center}
  {\tt (power (lcoset x h g) (ord x g) (quotient g h))} = {\tt (lcoset (e g) h g)}
\end{center}
where
\begin{center}
  {\tt (lcoset (e g) h g)} = {\tt (e (quotient g h))}
\end{center}
By {\tt divides-ord}, {\tt (ord x g)} is divisible by {\tt (ord (lcoset x h g) (quotient g h))}.  Therefore, if {\tt g} has
no element of order divisible by an integer {\tt m}, then neither does {\tt (quotient g h)}:
\begin{small}
\begin{verbatim}
  (defthm lift-elt-of-ord
    (implies (and (normalp h g)
                  (posp m)
                  (elt-of-ord m (quotient g h)))
             (elt-of-ord m g)))
\end{verbatim}
\end{small}
Now assume that {\tt g} is abelian.  Then every subgroup of {\tt g} is abelian and normal.  Since the quotient group of any nontrivial
cyclic subgroup of {\tt g} has a smaller order than {\tt g}, we have our induction scheme:
\begin{small}
\begin{verbatim}
  (defun cauchy-induction (g)
    (declare (xargs :measure (order g)))
    (if (and (groupp g)
             (abelianp g)
             (> (order g) 1))
        (cauchy-induction (quotient g (cyclic (cadr (elts g)) g)))
      ()))

  (defthm cauchy-abelian-lemma
    (implies (and (groupp g)
                  (abelianp g)
                  (primep p)
                  (divides p (order g)))
             (elt-of-ord p g))
    :hints (("Goal" :induct (cauchy-induction g))))
\end{verbatim}
\end{small}
In the proof of the above lemma, ACL2 generates a single nontrivial subgoal, the hypothesis of which is the instantiation of the goal
with
\begin{center}
{\tt (quotient g (cyclic (cadr (elts g)) g))}
\end{center}
substituted for {\tt g}.  By {\tt divides-order-quotient}, the order of this quotient group is divisible by {\tt p}, and therefore, by hypothesis,
it has an element of order {\tt p}.  The subgoal follows from {\tt lift-elt-of-ord}.

The final theorem is an immediate consequence of {\tt cauchy-abelian-lemma} and {\tt elt-of-ord-ord}:
\begin{small}
\begin{verbatim}
  (defthm cauchy-abelian
    (implies (and (groupp g)
                  (abelianp g)
                  (primep p)
                  (divides p (order g)))
             (and (in (elt-of-ord p g) g)
                  (equal (ord (elt-of-ord p g) g) p))))
\end{verbatim}
\end{small}

\begin{small}
\begin{verbatim}
\end{verbatim}
\end{small}

\section{Conjugacy Classes and the Class Equation}

The general case of Cauchy's Theorem ia based on an expression for the order of a group derived from a partition of its elements into
{\it conjugacy classes}.  The orsered list of conjugates of {\tt x} is {\tt (conjs x g)}, computed by
\begin{small}
\begin{verbatim}
  (defun conjs-aux (x l g)
    (if (consp l)
        (if (member-equal (conj x (car l) g)
                          (conjs-aux x (cdr l) g))
	    (conjs-aux x (cdr l) g)
          (insert (conj x (car l) g)
	          (conjs-aux x (cdr l) g)
                  g))
      ()))
  (defund conjs (x g) (conjs-aux x (elts g) g))
\end{verbatim}
\end{small}
Conjugacy is easily shown to be an equivalence relation, and it follows that intersecting classes are equal:
\begin{small}
\begin{verbatim}
  (defthmd equal-conjs
    (implies (and (groupp g) (in x g) (in y g)
                  (member-equal y (conjs x g)))
             (equal (conjs y g) (conjs x g))))
\end{verbatim}
\end{small}
We define a bijection between the conjugates of {\tt x} and the cosets of its centralizer:
\begin{small}
\begin{verbatim}
  (defund conj2coset (y x g) (lcoset (inv (conjer y x g) g) (centralizer x g) g))
  (defund coset2conj (c x g) (conj x (inv (car c) g) g))
  (defthm coset2conj-conj2coset
    (implies (and (groupp g) (in x g) (member-equal y (conjs x g)))
             (equal (coset2conj (conj2coset y x g) x g)
                    y)))
  (defthm conj2coset-coset2conj
    (implies (and (groupp g) (in x g) (member-equal c (lcosets (centralizer x g) g)))
             (equal (conj2coset (coset2conj c x g) x g)
                    c)))
\end{verbatim}
\end{small}
It follows that the size of the conjugacy class is the index of the centralizer:
\begin{small}
\begin{verbatim}
  (defthm len-conjs-cosets
  (implies (and (groupp g) (in x g))
           (equal (len (conjs x g))
                  (subgroup-index (centralizer x g) g))))
\end{verbatim}
\end{small}
Since {\tt (len (conjs x g))} = 1 iff {\tt (in x (center g))}, a list of the nontrivial conjugacy classes is computed by
\begin{small}
\begin{verbatim}
  (defun conjs-list-aux (l g)
    (if (consp l)
        (let ((conjs (conjs-list-aux (cdr l) g)))
          (if (or (in (car l) (center g))
                  (member-list (car l) conjs))
              conjs
            (cons (conjs (car l) g) conjs)))
      ()))
  (defund conjs-list (g) (conjs-list-aux (elts g) g))
\end{verbatim}
\end{small}
Thus, we can show that the following is a list of distinct elements that contains every element of {\tt g}:
\begin{small}
\begin{verbatim}
  (append (elts (center g)) (append-list (conjs-list g))))
\end{verbatim}
\end{small}
As a consequence, we have the {\it class equation}:
\begin{small}
\begin{verbatim}
  (defthmd class-equation
    (implies (groupp g)
             (equal (len (append (elts (center g)) (append-list (conjs-list g))))
                    (order g))))
\end{verbatim}
\end{small}

\section{General Case of Cauchy's Theorem}

Assume that the order of {\tt g} is divisible by a prime {\tt p}.  The function {\tt find-elt} searches for an element outside the center of
{\tt g} that has a centralizer with order divisible by {\tt p}:
\begin{small}
\begin{verbatim}
  (defun find-elt-aux (l g p)
    (if (consp l)
        (if (and (not (in (car l) (center g)))
                 (divides p (order (centralizer (car l) g))))
            (car l)
          (find-elt-aux (cdr l) g p))
      ()))
  (defund find-elt (g p) (find-elt-aux (elts g) g p))
\end{verbatim}
\end{small}
If such an element exists, then since it is not in the center, the order of its centralizer is less than that of {\tt g}.  This observation
provides our induction scheme:
\begin{small}
\begin{verbatim}
(defun cauchy-induction (g p)
  (declare (xargs :measure (order g)))
  (if (and (groupp g) (primep p) (find-elt g p))
      (cauchy-induction (centralizer (find-elt g p) g) p)
    t))
\end{verbatim}
\end{small}
On the other hand, if no such element exists, then for every non-central element {\tt x}, {\tt lagrange} implies that the index of the
centralizer of {\tt x} is divisible by {\tt p}, and by {\tt len-conjs-cosets}, so is {\tt (len (conjs x g))}.  According to the class
equation, the same is true of {\tt (center g)}.  Since {\tt (center g)} is abelian, we may apply {\tt cauchy-abelian} to complete the induction,
and we have

\begin{small}
\begin{verbatim}
  (defthmd cauchy
    (implies (and (groupp g) (primep p) (divides p (order g)))
             (and (in (elt-of-ord p g) g)
                  (equal (ord (elt-of-ord p g) g) p))))
\end{verbatim}
\end{small}

\section{Conclusion}

In 2007, Georges Gonthier et al.~\cite{georges} embarked on a formalization of finite group theory in Coq, with the objective of a machine-checked
proof of the Feit-Thompson Theorem: {\it All groups of odd order are solvable.}  Six years later, the ultimate success of this undertaking was
announced in an Inria Technical Report \cite{gonthier} listing fifteen coauthors.  In light of the experience of our present project, it would be
unsurprising to find that the Inria result did indeed involve some ninety man-years of effort.

The leap in complexity from Cauchy's Theorem to Feit-Thompson is daunting, but as C.~S.~Lewis reminds us, “With the possible exception of the
equator, everything begins somewhere.”  We may find further solace in a plan to pursue a direction orthogonal to Inria's objective of the
classification of finite groups, and better suited to ACL2's strengths.  While not specifically designed for the formalization of higher mathematics,
ACL2 is equipped with sophisticated procedures for managing rational arithmetic and polynomials. \cite{krug}  Algebraic number theory, the study of
finite extension fields of the rationals and their Galois groups, could be a fruitful application area founded on a formalization of elementary
group theory.  We have already demonstrated that our approach provides group computations in a straightforward manner, and we may anticipate that
the computational power and proof automation of ACL2 can be brought to bear on the analysis and verification of a variety of number-theoretic
algorithms of practical significance.  Clearly, there is much work to be done before such a plan can be more than a fanciful dream.

\nocite{*}
\bibstyle{eptcs}
\bibliographystyle{eptcs}
\bibliography{groups}
\end{document}